\documentclass[12pt]{article}
\usepackage{amsmath,amssymb,amsfonts,amscd,mathrsfs}
\usepackage{xcolor}
\definecolor{darkblue}{rgb}{0.1,0.1,.7}
\usepackage[colorlinks, linkcolor=darkblue, citecolor=darkblue, urlcolor=darkblue, linktocpage]{hyperref} 
\usepackage[square, comma, compress,numbers]{natbib}
\usepackage[]{graphicx}
\usepackage{geometry}
\usepackage[margin=10pt,font=small,labelfont=bf]{caption}
\usepackage{ifthen}

\definecolor{purp}{rgb}{0.6,0.1,.7}

\newcommand{\mrm}[1]{{\mathrm #1}}

\newcommand{\DD}[0]{\Delta}

\def\l{\ell} % can be switched to \ell if wanted 

\def\la{\lambda}

\newcommand{\reef}[1]{(\ref{#1})}

\newcommand{\vareps}{\varepsilon}
\newcommand{\eps}{\epsilon}
\def\beq{\begin{equation}} 
\def\eeq{\end{equation}} 
\def\del {\partial} 
\def\nn{\nonumber} 
\def\bZ {\mathbb{Z}} 
 
\def\bC {\mathbb{C}} 
\def\calO {{\cal O}} 
 
\def\calD {{\cal D}}

\def\gl{\lambda}

\def\bZ {\mathbb{Z}} 
\def\half{\textstyle\frac 12}
\def\quarter{\textstyle\frac 14}
\def\ge{\geqslant}
\def\le{\leqslant}

\newcommand{\diffop}[2]{\ifthenelse{\equal{#2}{1}}{\frac{\mrm{d}}{\mrm{d} #1}}{\frac{\mrm{d}^#2}{\mrm{d} #1^#2}}}

\numberwithin{equation}{section}
\def \d{{\rm d}}

\def \d{{\rm d}}

\def \vep{\varepsilon}
\def \half{{\textstyle {1 \over 2}}}

\def \quar{{\textstyle {1 \over 4}}}
\def \ts{\textstyle}

\begin{document}

\vspace*{-.6in} \thispagestyle{empty}
\begin{flushright}
DAMTP--2013-16\\
LPTENS--13/10\\
CERN-PH-TH/2013-092
\end{flushright}
\vspace{1cm} {\Large
\begin{center}
{\bf Diagonal Limit for Conformal Blocks in $d$ Dimensions}\\
\end{center}}
\vspace{1cm}
\begin{center}
{\bf Matthijs Hogervorst$^{a,b}$, Hugh Osborn$^{c}$, Slava Rychkov$^{b,a,d}$}\\[2cm] 
{
$^{a}$ Laboratoire de Physique Th\'{e}orique de l'\'{E}cole Normale Sup\'{e}rieure, Paris, France\\
$^{b}$ CERN, Theory Division, Geneva, Switzerland\\
$^{c}$ Department of Applied Mathematics and Theoretical Physics, Cambridge, England\\
$^{d}$ Facult\'{e} de Physique, Universit\'{e} Pierre et Marie Curie, Paris, France}

\end{center}

\vspace{4mm}

\begin{abstract}
Conformal blocks in any number of dimensions depend on two variables $z$, $\bar z$. Here we study their restriction to the special ``diagonal" kinematics $z=\bar z$, previously found useful as a starting point for the conformal bootstrap analysis. We show that conformal blocks on the diagonal satisfy ordinary differential equations, third-order for spin zero and fourth-order for the general case. These ODEs determine the blocks uniquely and lead to an efficient numerical evaluation algorithm. For equal external operator dimensions, we find closed-form solutions in terms of finite sums of ${}_3F_2$ functions. 
\end{abstract}
\vspace{.2in}
\vspace{.3in}
%\vskip 1cm 
\hspace{0.7cm} May 2013

\newpage

\tableofcontents 

\section{Introduction}
\label{sec:intro}
Many interesting results about Conformal Field Theory (CFT) in dimensions $d\ge3$ have been obtained recently using the bootstrap approach~\cite{Rattazzi:2008pe,Rychkov:2009ij,Heemskerk:2009pn,Caracciolo:2009bx,Poland:2010wg,Rattazzi:2010gj, Rattazzi:2010yc,Vichi:2011ux,Poland:2011ey,ElShowk:2012ht,Liendo:2012hy,ElShowk:2012hu,Fitzpatrick:2012yx,Komargodski:2012ek,Beem:2013qxa}. 
Conformal blocks,\footnote{In CFT${}_2$, one distinguishes the ``big" conformal blocks defined summing over the Virasoro descendants, and the ``small" blocks defined summing over the $SL(2,\bC)$ descendants only. The CFT$_d$ blocks considered here reduce to the ``small" blocks for $d=2$.} which sum up the contributions of a primary operator and its descendants in the operator product expansion,  play a fundamental role in this approach, and their detailed knowledge is a prerequisite for any advance.
Although the theory of these objects was initiated in the 70's \cite{Ferrara:1973vz,Ferrara:1974nf,Ferrara:1974ny,Polyakov:1974gs}, and much developed recently \cite{DO1,DO2,DO3,Costa:2011dw,ElShowk:2012ht,SimmonsDuffin:2012uy,Osborn:2012vt,Hogervorst:2013sma,Fitzpatrick:2013sya}, it is still incomplete. 

To introduce some notation and formulate our problem, consider a correlation function of four scalar primary\footnote{These are called quasi-primaries in CFT${}_2$.} operators $\phi_i$ of dimension $\Delta_i$. Conformal invariance restricts this correlator to have the form 
\begin{equation}
\langle\phi_{1}(x_{1})\phi_{2}(x_{2})\phi_{3}(x_{3})\phi_{4}(x_{4})\rangle
=\left(  \frac{x^2_{24}}{x^2_{14}}\right)  ^{\frac12\Delta_{12}%
}\left(  \frac{x^2_{14}}{x^2_{13}}\right)  ^{\frac12\Delta_{34}}
\frac{g(u,v)}{(x^2_{12})^{\frac12(\Delta_{1}+\Delta_{2})}(x_{34}^2)^{\frac12(\Delta_{3}+\Delta_{4})%
}}\,, 
\label{eq:4pt}%
\end{equation}
where $x_{ij}\equiv x_i-x_j$, $\Delta_{ij}\equiv \Delta_i-\Delta_j$, and $g(u,v)$ is a function of the conformally invariant cross-ratios
\begin{equation}
u=\frac{{x_{12}^{2}\, x_{34}^{2}}}{{x_{13}^{2}\, x_{24}^{2}}}\, ,\qquad 
v=\frac{{x_{14}^{2}\, x_{23}^{2}}}{{x_{13}^{2}\, x_{24}^{2}}}\,.
\end{equation}

The four point function \reef{eq:4pt} can be expanded into conformal partial waves corresponding to the primaries $\calO$ appearing 
in the operator product expansions (OPEs) $\phi_1\times\phi_2$ and $\phi_3\times\phi_4$. Each conformal partial wave has the same functional form as the four point function itself.\footnote{See e.g.~\cite{Costa:2011dw} for an explanation.} This gives the following series representation for the function $g(u,v)$:
\beq
g(u,v)= \sum_\calO f_{12\calO}f_{34\calO}\, {G_{\Delta,\l}(u,v)}\,,
\label{eq:CBexp}
\eeq
Here the $G_{\Delta,\l}(u,v)$ are the universal parts of the conformal partial waves; these are the conformal blocks mentioned above. They depend on the dimension $\Delta$ and spin $\l$ of the exchanged primary $\calO$, as well as on the external dimension differences $\Delta_{12}$ and $\Delta_{34}$. The $f_{ij\calO}$ are the OPE coefficients which depend on the CFT under consideration. 
The representation \reef{eq:CBexp} has a nonzero radius of convergence \cite{Pappadopulo:2012jk} and is the starting point for the conformal bootstrap studies cited above.

An equivalent representation of the conformal blocks is expressed in terms of 
the symmetric functions of  variables $z, \bar{z}$ which are related to 
$u,v$ via
\begin{equation}
u=z\bar{z},\qquad v=(1-z)(1-\bar{z})\,.
 \label{eq:uvzzbar}%
\end{equation}
Then
\begin{equation}
{G_{\Delta,\l}(u,v)} = F_{\lambda_1 \lambda_2} ( z, {\bar z}) \, ,
\end{equation}
for 
\beq
\Delta=\lambda_1+\lambda_2,\quad \l=\lambda_1-\lambda_2\in\{0,1,2,\ldots\}\,.
\eeq
The normalisation of conformal blocks is a matter of choice, for us
\beq
 F_{\lambda_1 \lambda_2} ( z, z)  \sim z^\Delta \quad \mbox{as} \quad z\to 0 \, .
\label{norm0}
\eeq

For even $d$, conformal blocks have closed-form expressions in terms of hypergeometric functions \cite{Ferrara:1974ny,DO1,DO2,DO3}, which proved very useful for conformal bootstrap applications. For example, the $4d$ blocks are given by \cite{DO1}
\begin{gather}
 F_{\lambda_1 \lambda_2} ( z, {\bar z})\big |_{d=4}  =\frac{1}{\l+1}
\frac{z\bar{z}}{z-\bar{z}}\big [ k_{2\lambda_1}(z)k_{2\lambda_2 -2}
(\bar{z})-(z\leftrightarrow\bar{z})\big ]\,,
\label{eq:DO}\\[2pt]
k_{\beta}(z)  = z^{\beta/2}{}_{2}F_{1}\left(  \half(\beta-\Delta_{12}),\half(\beta+\Delta_{34});\beta;z\right)\,, \nn
\end{gather}
For general $d$, a numerical approach to computing conformal blocks was developed in 2012 in \cite{ElShowk:2012ht}, where it was applied in the bootstrap analysis of the 3$d$ Ising model. First, one computes the blocks on the ``diagonal" $z=\bar z$, which turns out to be significantly simpler than for general $z,\bar z$. Then, one computes derivatives in the direction normal to the diagonal, using recursion relations following from a second-order partial differential equation (PDE) that the blocks satisfy. While this method cannot compute the blocks at a finite distance from the diagonal,\footnote{See \cite{DO2,Hogervorst:2013sma} for a proposal on how to do this, using power series expansions around $z,\bar z=0$.}  it proved very efficient in computing them in the derivative expansion around $z=\bar z=\frac{1}{2}$. 
The latter information is sufficient for applying the existing conformal bootstrap algorithms.

Motivated by \cite{ElShowk:2012ht}, here we analyse in more detail the diagonal limit of conformal blocks.
The central part of the paper is section \ref{sec:diffeq}, where we systematically derive ordinary differential equations (ODEs) satisfied on the diagonal.
While it is well known that conformal blocks as functions of $z,\bar z$ satisfy PDEs (which arise from eigenvalue equations for the Casimir operator), it is by no means evident that the diagonal limit satisfies an equation by itself. As we will see, this follows from an interplay between the well-known quadratic and the rarely-used quartic Casimir PDEs.

In the rest of the paper we discuss various applications of the ODEs from section \ref{sec:diffeq}. In section \ref{sec:a=0} we show that setting one pair of the external dimensions equal, the scalar conformal block ODE can be integrated in terms of a ${}_3F_2$ function, ``explaining" the results of \cite{ElShowk:2012ht,Paulos-notes}. For the same case $\Delta_{1}=\Delta_2$, we also find succinct expressions of higher spin blocks as finite sums of ${}_3F_2$ functions.

In section \ref{sec:algo} we propose to use the ODEs from section \ref{sec:diffeq} as a basis for an efficient algorithm which can numerically compute conformal blocks and their derivatives around any point on the diagonal. This algorithm is a generalisation and an improvement of the method used in \cite{ElShowk:2012ht} for the case of equal external scalar dimensions. 

We summarise the main aspects of this paper in section \ref{sec:summary}. Appendices \ref{sec:fact} and \ref{sec:recursions} discuss how our ODEs can be used to generate power series expansions of the conformal blocks, and a few special cases when the ODEs factorise so that their order can be reduced.

\section{Differential equations on the diagonal}
\label{sec:diffeq}

Conformal blocks in CFT${}_{d}$ may be extended to define symmetric analytic 
functions of two complex variables $z$ and $\bar z$, everywhere in $\bC^2$ except, as follows
from \reef{eq:DO} for $d=4$, for branch points at $z,\bar z=0$,$1$ and $\infty$. For
Euclidean spacetime signature, it is necessary to impose $\bar z=z^*$ whereas for
a Minkowski signature $z,{\bar z}$ are two independent real variables.
%and conformal blocks can be viewed as smooth, real, real-analytic function of two real variables $x={\rm Re}\,z$ and $y={\rm Im}\,z$. Although we will not use complex analysis, we will keep $z$ as a convenient way of parametrizing the $(x,y)$ plane.

In this paper we study conformal blocks on the ``diagonal" $z=\bar z$. For the Euclidean
section $z=x+iy,  \,{\bar z} = x-iy$, for $x,y$ real, the limit $y\to 0$ is in general singular
for $x\in [1,\infty)$\footnote{This is also apparent from the four-dimensional result \eqref{eq:DO}
which can be expressed in the form $\frac{f(z) \, g({\bar z}) - f({\bar z}) \, g(z)}{z-{\bar z}}$
where $f(z),g(z)$ have branch cuts along the real axis for $ z>1$. Hence for $x>1$, $f(z),f({\bar z})$ and
$g(z),g({\bar z})$ approach different limits as $y\to 0$ and the denominator is not cancelled.}.
However away from the cut, for arbitrary complex $z$, the  limit ${\bar z}\to z$ is well defined and
conformal blocks on the diagonal will be denoted by
\beq
f_{\lambda_1\lambda_2}(z) = F_{\lambda_1\lambda_2}(z,z) \,.
\label{fdiag}
\eeq
For most conformal bootstrap applications we require real $z$ and $0<z<1$, 
which includes the crossing symmetric point $z=\bar z=\half$.
 
These functions allow a power series expansion of the form
\beq
\label{eq:seriesf}
f_{\lambda_1\lambda_2}(z)=z^\Delta \sum_{n=0}^\infty a_n z^n\,,\qquad a_0=1\,, 
\eeq
where $a_0=1$ follows from the normalisation condition \eqref{norm0}.\footnote{This 
is the same normalisation as in \cite{DO3,ElShowk:2012ht,Hogervorst:2013sma}. 
It is related to the normalisation in \cite{DO1,DO2} by $G^{\text{here}}_{\Delta,\l}=(-2)^\l (\vareps)_\l/(2\vareps)_\l G^{\text{there}}_{\Delta,\l}$.}
Basic CFT properties imply $ z^{-\Delta} f_{\lambda_1\lambda_2}(z)$  is analytic in $z$ with singularities
at $z=1,\infty$ and so this expansion is convergent for $|z|<1$ \cite{Pappadopulo:2012jk,Hogervorst:2013sma}. 

As functions of $z$ and $\bar z$, conformal blocks satisfy various PDEs, derived by acting on the four point function with the Casimir operators of the conformal group and demanding that conformal partial waves be eigenfunctions. 
For general $d$, there are quadratic and quartic Casimir operators.
The corresponding second- \cite{DO2} and fourth-order \cite{DO3} PDEs are 
\beq
\Delta_2\, F_{\lambda_1\lambda_2}=c_2\,  F_{\lambda_1\lambda_2}\,,\qquad  
\DD_4 \, F_{\la_1\la_2}= c_4\,  F_{\la_1\la_2}\,,
\label{eq:24}
\eeq
where $(\vep=d/2-1)$
\begin{align}
c_2&= \lambda_1 ( \lambda_1  - 1 ) + \lambda_2 ( \lambda_2  - 1 - 2\vep )
%\equiv \half [(\lambda_1+\gl_2)(\gl_1+\gl_2 -2 -2\vep)+(\la_1 - \la_2)(\la_1 - \la_2 + 2\vareps)]\,,\nn\\
= \half [\l (\l+ 2\vareps)+\Delta(\Delta -2 -2\vep)]\,,\nn\\
c_4&= %(\la_1 - \la_2)(\la_1 - \la_2 + 2\vareps)(\la_1 + \la_2 - 1)(\la_1 + \la_2 -1-2\vareps)\,,
\l (\l + 2\vareps)(\Delta - 1)(\Delta -1-2\vareps)\,,
\label{Casimir}
\end{align}
and the differential operators can be written as
\begin{align}
\label{eq:CasOp} 
\DD_2 &= D_z + D_{\bar z} + 2\vareps \frac{z \bar z}{z-\bar z}
 \Big[(1-z) \diffop{z}{1} - (1-\bar z) \diffop{\bar z}{1} \Big]\,,\\
\label{eq:quartic} \DD_4 &= \left(\frac{z \bar z }{z-\bar z} \right)^{2\vareps} 
(D_z - D_{\bar z}) \left(\frac{z \bar z }{z-\bar z} \right)^{-2\vareps} (D_z - D_{\bar z})\,,
\end{align}
in terms of the one-dimensional differential operator of the ${}_2F_{1}$ type:
\beq
\label{hyper}
D_z\equiv D_z(a,b)= (1-z)z^2 \diffop{z}{2} -(a+b+1)\, z^2 \diffop{z}{1} - ab\, z\,.
\eeq
Here and below we denote $a=-\half \Delta_{12}$, $b=\half\Delta_{34}$.

For $\l=0$ the quartic eigenvalue $c_4$ vanishes. In this case the quartic Casimir 
equation actually reduces to a simple second-order PDE \cite{DO3}
\begin{align}
\label{lzer}
\big ( D_z - D_{\bar z} \big ) F_{\lambda\,\lambda} = 0 \, .
\end{align}

We will now reduce the PDEs (\ref{eq:24},\ref{lzer}) to the diagonal. Conformal 
blocks are symmetric functions of $z,\bar z$ and near the diagonal \eqref{fdiag}
can be extended to a power series expansion of the form:
\begin{align}
F_{\lambda_1\lambda_2}(z,\bar z) = f_{\lambda_1\lambda_2}(t) 
+\quar (z-\bar z)^2 g_{\lambda_1\lambda_2}(t) + {\rm O}\big ( (z-\bar z)^4 \big ),
\quad t=\half(z+\bar z)\,.
\label{eq:neardiag}
\end{align}
The quadratic Casimir equation at $z=\bar z$ gives one relation between 
$  f \equiv f_{\lambda_1\lambda_2}$ and $g\equiv g_{\lambda_1\lambda_2}$:
\beq
\bigg [\half (1-z) z^2 \diffop{z}{2}-(1+a+b+\vareps ) z^2 \diffop{z}{1}
 -2 a b \, z-c_2 \bigg ]f(z) +(1+2\vareps)(1-z)z^2 g(z)= 0\,.
\label{eq:2diag}
\eeq

A second relation follows from the quartic Casimir equation, which reduces 
to a third-order PDE on the diagonal, so that the ${\rm O}\big ( (z-\bar z)^4 \big )$ 
terms omitted in \reef{eq:neardiag} do not contribute. This relation takes the 
schematic form
\beq
P_3({\rm d}_z) f(z)+ P_2({\rm d}_z) g(z)=0\,.
\label{eq:4diag}
\eeq
In addition, for $\l=0$, substituting \reef{eq:neardiag} into (\ref{lzer}) and looking 
at the ${\rm O}(z-\bar z)$ terms gives an equation of the form
\beq
Q_2({\rm d}_z) f(z)+ Q_1({\rm d}_z) g(z)=0\,,
\label{eq:scdiag}
\eeq

The $P_i$ and $Q_i$ in the last two equations are certain differential operators 
of degree $i$ with polynomial coefficients, whose precise form is important for 
what follows but is of no particular interest to write them down explicitly.

With the help of \reef{eq:2diag}, we can eliminate $g(z)$ from 
\reef{eq:4diag} and \reef{eq:scdiag}. This gives ODEs for $f(z)$ by itself, 
fourth-order for the general case, and third-order for $\l=0$: 
\begin{subequations}
\label{eq:ODEs}
\begin{align}
\label{eq:ODE4}
&D_z^{(4,3)}f_{\lambda_1\lambda_2}(z)=0\,, \\
\label{eq:ODE3}
&D_z^{(3,2)} f_{\lambda\, \lambda}(z)=0\,,
\end{align}
\end{subequations}
where in general $D^{(n,j)}$ are differential operators of the form
\beq 
\label{eq:Dnj}
D_z^{(n,j)} = (z-1)^j \, z^n {{\rm d}^n \over {\rm d
} z^n} + \sum_{r=n-j+1}^{n-1} (z-1)^{j-n+r} p_{r}(z) \, z^r {{\rm d}^r \over
{\rm d } z^r} + \sum_{r=0}^{n-j} p_{r}(z) \, z^r {{\rm d}^r \over {\rm d }
z^r}\,, 
\eeq 
for $p_{r}(z)$ polynomials of degree
\beq 
\deg p_r(z) = 
    \begin{cases}
       n-r , & r>n-j \, , \\
       j ,   & r \le n-j \,.
     \end{cases} \label{eq:cases}
\eeq
The differential operators are symmetric functions of $a,b$ and so, with
$P = 2ab$, $S = a+b$, the operator in \eqref{eq:ODE4} is determined by
\begin{align}
p_3(z) = {} & (4 S-2 \vareps +7)z +4 \vareps -2\, , \nn \\[4pt]
p_2(z) = {} & [2 P+(S-\vareps +2) (5 S-\vareps +5)] z^2  + 
2 \, [{c_2}-P+(3 S-\vareps +3) (2 \vareps -1)] z \nn \\
&{} +4 \vareps ^2-2\vareps -2{c_2} \, , \nn \\[4pt]
p_1(z) = {}& [4 P+(2 S+1) (S-\vareps +2)] (S-\vareps +1) z^3  \nn\\
& {}+[{c_2} (4 S-2 \vareps +1)+2 (2 S+1) (S-\vareps +1) (2 \vareps -1)+P (-4 S+10 
\vareps -5)] z^2 \nn\\
&{} + [-6 \vareps  P+P+2 (2 S+1) \vareps  (2 \vareps -1)+{c_2} (-4 S+6 \vareps +1)] 
z-2c_2 (2 \vareps  +1) \, ,  \nn\\[4pt]
p_0(z) ={} &2 P (S-\vareps ) (S-\vareps +1) z^3+(S-\vareps ) [{c_2} (2 S-1)+P (6 \vareps -1)] z^2\nn  \\
&{} + [{c_4}+2 (2 \vareps +1) ({c_2} (S-1)+P \vareps )] z-{c_4}+2{c_2} (2 \vareps +1)\,,
\label{D4p}
\end{align}
with $c_2,c_4$ as in \eqref{Casimir}.
In \eqref{eq:ODE3}
\begin{align}
  p_2(z) = {}& (3 S-\vareps +3)z  +2 \vareps \, , \nn\\[4pt]
  %% corrected:
  p_1(z) = {}& [2 P+(2 S+1) (S-\vareps +1)] z^2+2 ({c_2}-P+ 2 S \vareps +\vareps ) z-2 {c_2}  \, , \nn\\[4pt]
  p_0(z) = {}& 2 P (S-\vareps ) z^2+[ P(2\vareps+1)+{c_2} (2 S-1)] z+2 {c_2}\,, 
  \label{D3p}
\end{align}
where here $c_2= 2\lambda(\lambda -1 - \vareps)$.

With these expressions for the differential operators,
the ODEs in \eqref{eq:ODEs} are a crucial result for this paper. They allow
for a direct analysis of the diagonal limit of conformal blocks.
The differential operators satisfy the symmetry relations
\begin{align}
(1-z)^{2b-2} D_z^{(4,3)}(1-z)^{-2b} = {}& D_{z'}^{(4,3)} \big |_{a\to -a} \, , \nn \\
(1-z)^{2b-1} D_z^{(3,2)}(1-z)^{-2b} = {}& D_{z'}^{(3,2)} \big |_{a\to -a} \, ,
\label{DDprime}
\end{align}
for
\beq
z' = \frac{z}{z-1} \, ,
\label{zprime}
\eeq
and similarly for $a\leftrightarrow b$.

Differential operators of the form \eqref{eq:Dnj} subject to \eqref{eq:cases} with $j\le n$
have regular singular points at $0,\infty,1$ and so are Fuchsian. It is
straightforward to see that  $D^{(n,j)}$ has  $n(j+1) - {1\over 2} j(j-1)$ possible  independent parameters.
Such differential operators are characterised in part by their exponents, which are defined
by the roots of the indicial equation for series solutions about $0,\infty,1$. Thus if
$D_z^{(n,j)} u(z) = 0$ then
\beq
u \sim z^{s_{0,r}}  , \  z\to 0 \,, \quad u \sim z^{-s_{\infty,r}}  , \ z\to \infty  , \quad 
u \sim (z-1)^{s_{1,r}}  , \  z\to 1 \, , \quad r=1,\dots, n \, ,
\eeq
subject to the Fuchs relation
\beq
\sum_{r=1}^n ( s_{0,r} + s_{\infty,r} + s_{1,r} ) = \tfrac{1}{2} n (n-1) \, ,
\label{Frob}
\eeq
so that there are a  maximum of $3n-1$ independent exponents (although some may 
be degenerate and there may then be logarithmic  singularities).
For differential operators of the form \eqref{eq:Dnj} with \eqref{eq:cases}
and $j<n$ then the exponents for $z=1$ are constrained so that
we may take $s_{1,r} = r-1$, $r=1,\dots, n-j$.
$D_z^{(n,1)} u(z) = 0$  has solutions of the form $z^\lambda {}_nF_{n-1}(z)$
with $2n$ parameters.

For the differential operators $D^{(4,3)}$ and $D^{(3,2)}$ of interest here
it is easy to determine the exponents giving respectively, in terms of $\Delta,\ell,a,b,\vareps$,
\begin{subequations}
\label{eq:exponents}
\beq
\text{
\begin{tabular}{  c   c   c   }
0 &$ \infty$& 1 \\
\hline
$\Delta$ & $2a$ & 0 \\ 
$2+ 2 \vareps - \Delta$& $2b$ &$ -2a-2b$ \\
$\ell + 1 + 2\vareps$ & $a+b -\vareps$ & $- a -b -\vareps $\\
$1-\ell$ & $1+a+b - \vareps$ &  $1-a-b - \vareps $
\end{tabular}
}
\eeq
and, with $\ell=0$,
\beq
\text{
\begin{tabular}{  c   c   c   }
0 &$ \infty$& 1 \\
\hline
$\Delta$ & $2a$ & 0 \\ 
$2+ 2 \vareps - \Delta$& $2b$ &$ -2a-2b$ \\
$1$ & $a+b -\vareps$ & $- a -b -\vareps $\\
\end{tabular}
}
\eeq
\end{subequations}
Of course \eqref{Frob} is satisfied in both cases. The detailed exponents are useful
later in discussing possible factorisations of $D_z^{(4,3)}$ and $D_z^{(3,2)}$ in
special cases (see appendix \ref{sec:fact}).

For general $a,b$ there is no apparent  solution of either \eqref{eq:ODE4} or more simply
\eqref{eq:ODE3} in terms of known special functions, although
results for $a=0$ are obtained in section \ref{sec:a=0}. However, solutions as  a power series expansion 
around any of the regular singular points are generated by Frobenius' method. We are interested in expansions around $z=0$
with leading term $z^\Delta$.
Frobenius' method gives recursion relations by which all the coefficients $a_n$ in \reef{eq:seriesf} can be determined from $a_0=1$. These recursions, four-term for the general case and three-term for $\l=0$, are given in appendix~\ref{sec:recursions}. In section \ref{sec:algo} we will explain how these and related recursions can be used to efficiently evaluate conformal blocks in the derivative expansion around any point on the diagonal (and in particular the point $z=\bar z=\frac{1}{2}$ relevant for the conformal bootstrap applications).

Mathematically, it is natural to extend the definition of $f_{\lambda_1\,\lambda_2}$ 
to arbitrary real $\lambda_1,\lambda_2$ (i.e. $\l$ not necessarily a non-negative integer) 
as solutions of \reef{eq:ODE4}. In this case the transformations 
\begin{subequations}
\begin{align}
\label{eq:Dtrans}\Delta \to {}& 2+2\vareps - \Delta,\qquad \l \to \l \,,\\
\label{eq:ltrans}\ell \to{}&   - \ell  - 2 \vareps,\qquad \, \  \Delta \to \Delta\,,
\end{align}
\end{subequations}
under which both $c_2$ and $c_4$ are invariant, map solutions
into solutions. For 
physical dimensions $2\vareps \in {\mathbb N}$.
Conformal blocks,  with the leading term
 $z^{\lambda_1+\lambda_2}$, are related 
by \reef{eq:Dtrans} to solutions with leading term
 $z^{2 + 2\vareps - \lambda_1 -\lambda_2}$.
On the other hand conformal blocks 
with the same $\Delta$ and spins related by \reef{eq:ltrans} are identical:
\beq
\label{eq:HOsym}
f_{\lambda_1-\vep\,\lambda_2+\vep}(z)=f_{\lambda_1\,\lambda_2}(z)\, ,
\eeq
since they both satisfy the normalisation condition \eqref{eq:seriesf}.

\section{Closed-form results for $a = 0$}
\label{sec:a=0}
%\subsection{Scalar case}

It was found in Ref.~\cite{ElShowk:2012ht} that for $a = b = 0$ the functions
 $f_{\la\, \la}(z)$ and $f_{\la\, \la - 1}(z)$ can be expressed via generalised 
 hypergeometric functions ${}_3F_2.$ The purpose of this section is to 
 give a different derivation of that result, and extend it to higher $\l$. 
 We will also be able to generalise this solution to the case when only 
 one of the external dimension differences is required to be zero. 
 For definiteness we consider $a=0$ and  allow $b$ to be arbitrary. 

The essential idea is to consider solutions of the following form: 
\beq
\label{fFZ}
f(z) =  { 1\over (1-z)^b}\,  F (Z)  \quad \mbox{or}  \quad
f(z) =  { 1\over (1-z)^b} \bigg ( \frac{1}{z} - \frac12 \bigg ) \, F (Z) 
\eeq
for
\beq
 Z = z+z' = zz'={z^2 \over 4(z-1)} \, ,
\eeq
where $z'$ is given  by \eqref{zprime}. Since $\frac{1}{z} - \frac{1}{2} = \half(\frac{1}{z} - \frac{1}{z'})$,
\eqref{fFZ} requires that $(1-z)^b f(z)$ is even or odd under $z\leftrightarrow z'$.

The utility of the ansatz \eqref{fFZ} follows from crossing symmetry 
of the four point function under $x_1\leftrightarrow x_2$.
This implies that  conformal blocks satisfy the relation \cite{DO1}:
\beq
\label{eq:crossF}
F_{\gl_1\gl_2}(z',\bar z') = (-1)^\l (1-z)^{b} (1-\bar z)^{b} F_{\gl_1 \gl_2}(z,\bar z)\big |_{a\to -a} \, .
\eeq
On the diagonal the corresponding relation becomes:
\beq 
\label{eq:crossf}
f_{\gl_1\gl_2}(z') = e^{\pm i\pi \Delta} (1-z)^{2b} f_{\gl_1\gl_2}(z)\big |_{a\to -a} \,.
\eeq
The factor $e^{\pm i\pi \Delta}$ arises since $z\to z'$ maps $[0,1) \to (-\infty,0]$,
with $\pm$ according which side of the branch cut on the negative axis
$f_{\gl_1\gl_2}(z') $ is evaluated on. 

The reason why $a=0$ is special is that in this case both sides of \reef{eq:crossf}
 involve the same conformal block and therefore the transformation 
\beq
\label{eq:trans}
z\to z',\quad f(z)\to f(z')= (1-z)^{2b} f(z)
\eeq
should preserve the defining equations \reef{eq:ODEs}, as is
implied  by eqs.~\eqref{DDprime} when $a=0$.

In consequence writing
\beq
f_{\lambda\lambda}(z) = (1-z)^{-b} F_{\lambda\, 0}(Z) \, ,
\label{fzero}
\eeq
ensures,  since $(1-z)^{b}f_{\lambda\lambda}(z) $ has the same exponents
at $z=1,\infty$ and $z\to Z$  maps $1,\infty \to \infty$, that we may obtain
from \eqref{eq:ODE3} a corresponding differential equation for $F_{\lambda\, 0}(Z)$. 
This equation becomes 
\begin{align}
\label{Feig}
%{\D}^{(3,1)}_Z \hat f_{\lambda_1\lambda_2}(Z)=0\, ,\qquad 
{D}^{(3,1)}_Z F_{\lambda\, 0}(Z)=0\,,
\end{align}
 involving a simpler $D^{(3,1)}$ differential operator:
\begin{align}
\label{DZ}
{D}^{(3,1)}_Z &=  (Z-1) Z^3{\d^3\over \d Z^3} + \big ( {\ts {3\over 2}} ( 2 Z-1) 
 - \vep (Z-1) \big ) Z^2 {\d^2 \over \d Z^2} \cr 
&\qquad\qquad{} - \big ( (b^2-1+\vep ) Z - \half c_2 - \half \vep \big ) Z {\d \over \d Z} +  \vep b^2 Z - \quarter c_2  \\[4pt]
&=  Z (\delta_Z -\vep)(\delta_Z-b)(\delta_Z+b) \
- ( \delta_Z - \half )(\delta_Z - \lambda )(\delta_Z + \lambda - 1 -\vep ) \, ,
\label{D31}
\end{align}
for  $\delta_Z = Z \frac{\d}{\d Z}$.
The form \eqref{D31} leads naturally to a solution in terms  of  ${}_3 F_2$ functions, as 
mentioned already.
Hence the solution of \eqref{Feig}, consistent with the condition \eqref{norm0} at $Z=0$, is:
\begin{align}
\label{Fone}
F_{\lambda \, 0} (Z) = (-4Z)^{\lambda}
{}_3F_2\bigg ({\lambda-\vep,\lambda+b,\lambda-b\atop 2\lambda-\vep, \lambda+\half} ;
Z \bigg ) \,.
\end{align}
This result was obtained in \cite{ElShowk:2012ht} for $b=0$ (and generalised to nonzero 
$b$ in \cite{Paulos-notes}) by a different method,  involving 
directly manipulating a power series representation of the scalar block.

Applying the same procedure to \reef{eq:ODE4} leads  to an equation involving an operator ${D}_Z^{(4,2)}$.
 This is not manifestly  soluble in terms of known special functions.\footnote{An alternative 
 ${D}_Z^{(4,2)}$ equation can be obtained using \reef{fFZ} with an extra factor of $(1/z-1/2)$ on the RHS.}
 To find the diagonal conformal blocks for general $\l$ we use instead recursion relations,
 as discussed in the next section.

\subsection{Recursion relations for different $\ell$}

Ref.~\cite{DO3} derived differential recursion relations relating $F_{\lambda_1 \lambda_2}(z,\bar z)$ of different spin (and the same $a,b$).
It was noticed in \cite{ElShowk:2012ht} that these relations are particularly useful for relating the diagonal limits of conformal blocks, since many derivative terms vanish on the diagonal. Here we follow the same idea and set $z=\bar z$ in \cite{DO3}, eq.~(4.28), $i=0,1,2$. 
We get the following three recursion relations for the $f_{\lambda_1 \lambda_2}(z)$:
\begin{subequations}
\label{recurf}
\begin{align}
%rec1
&\bigg ({2\over z}-1 + {ab \, (c_{2} + 2 \vep )
\over 2\lambda_1 (\lambda_1-1)(\lambda_2-\vep)(\lambda_2-1-\vep)}  \bigg ) 
f_{\lambda_1 \lambda_2}(z)  
\nn\\
&\qquad{}= {\ell+2\vep\over \ell+\vep} \,\big ( f_{\lambda_1\, \lambda_2-1}(z) 
+ t_{\Delta} \, \beta_{\lambda_1} \,f_{\lambda_1+1\, \lambda_2}(z) 
\big ) 
\nn\\
& \qquad\qquad {}+ {\ell\over \ell+\vep} \, \big ( f_{\lambda_1-1\, \lambda_2}(z) 
+ t_{\Delta} \, \beta_{\lambda_2-\vep} \,
f_{\lambda_1\, \lambda_2+1}(z)   \big )  \, , \\
%rec2
&\bigg ((1-z){\d \over \d z} - a -b  + 
(1+\vep) \, {ab \, (c_{2} + 2 \vep )  
\over 2\lambda_1 (\lambda_1-1)(\lambda_2-\vep)(\lambda_2-1-\vep)}  \bigg ) 
f_{\lambda_1 \lambda_2}(z)  
\nn\\
&\qquad{}= {\ell+2\vep\over \ell+\vep} \,\big ( \lambda_2 \, 
f_{\lambda_1\, \lambda_2-1}(z) - t_{\Delta} \, \beta_{\lambda_1} \,
(\lambda_1-1-\vep) \, f_{\lambda_1+1\, \lambda_2}(z)   \big ) 
\nn\\
& \qquad\qquad {}+ {\ell\over \ell+\vep} \,\big ( (\lambda_1+\vep) \,
f_{\lambda_1-1\, \lambda_2}(z)      
- t_{\Delta} \, \beta_{\lambda_2-\vep} \, (\lambda_2-1-2\vep) \,
f_{\lambda_1\, \lambda_2+1}(z)   \big ) \, , 
\\
%rec3
&{} (\Delta-1) (\Delta-1-2\vep) \,  
{ab\,(\ell+\vep) 
\over 2\lambda_1 (\lambda_1-1)(\lambda_2-\vep)(\lambda_2-1-\vep)} \,
f_{\lambda_1 \lambda_2}(z) 
\nn\\
&\qquad =(\Delta-1) \big ( f_{\lambda_1\, \lambda_2-1}(z)  -
f_{\lambda_1-1\, \lambda_2}(z) \big ) 
\nn\\
&\qquad\qquad{} - (\Delta-1-2\vep) t_{\Delta} \, 
\big (  \beta_{\lambda_1} \, f_{\lambda_1 +1 \, \lambda_2}(z)  -
 \beta_{\lambda_2-\vep} \, f_{\lambda_1\, \lambda_2 + 1 }(z) \big ) \,,
\end{align}
\end{subequations}
where
\begin{align}
\label{deftb}
t_\Delta = {(\Delta-2\vep)(\Delta-1)\over (\Delta-\vep)(\Delta-1-\vep)} \, , \quad
\beta_\lambda = {(\lambda+a)(\lambda-a)(\lambda+b)(\lambda-b) \over
4\lambda^2(2\lambda-1)(2\lambda+1)} \, . 
\end{align}
These recurrence  relations respect the symmetry \eqref{eq:crossf} and also \reef{eq:HOsym}. They are potentially derivable directly
 from the 4th-order operator $D^{(4,3)}$ along the lines
of the approach in \cite{DO3} but we do not consider that here.

Eqs. \eqref{recurf}  may be combined in various ways to remove different terms.
Thus
\begin{align}
\label{recurfa}
&  {\Delta-1-2\vep \over \Delta-1- \vep } 
\bigg ((\Delta-2-2\vep) \Big (\, {1\over z}-{1\over 2}\, \Big )
+ (1-z){\d \over \d z} - a -b \cr
\noalign{\vskip -4pt}
& \hskip 6.5cm {} + {ab \, (\Delta-2 )
\over 2 (\lambda_1-1)(\lambda_2-1-\vep)}  \bigg )
f_{\lambda_1 \lambda_2}(z)  \cr
&{}= {1\over \ell + \vep} \,
\big ( (2\lambda_1 -1) \ell \, f_{\lambda_1-1\, \lambda_2}(z)
+ (2\lambda_2 -1 - 2\vep ) (\ell+2\vep)  \, f_{\lambda_1\, \lambda_2-1}(z)
\big ) \, ,
\end{align}
which may be used to determine $f_{\lambda\, \lambda- \ell}$ starting from
$\ell=0$.

\subsection{Solving the recursion relations}

For conformal blocks for $a=0$ and $\ell = 0,1,2, \dots$ we extend
\eqref{fzero} to define
\begin{align}
\label{fgZ}
f_{\lambda\,\lambda-2n}(z)\big |_{a=0} = {}& 
{1\over (1-z)^b} \, F_{\lambda \, 2n} (Z) \, , \cr
f_{\lambda\,\lambda-2n-1}(z)\big |_{a=0} = {}& {1\over (1-z)^b}
\Big ( \, {1\over z} - {1\over 2} \Big )  \, F_{\lambda\, 2n+1} (Z) \, .
\end{align}
The normalisation condition \reef{norm0} requires
\begin{align}
\label{fZlim}
F_{\lambda \, 2n} (Z)  , F_{\lambda\, 2n+1} (Z) \sim (-4Z)^{\lambda - n} \quad \hbox{as}
\quad Z \to 0 \, .
\end{align}

%For off diagonal results if
%\begin{ali
%\label{gGZ}
%g_{\lambda\lambda} (z) =  { 1\over (1-z)^b} \, G_{\lambda 0}(Z) \, , 
%\end{align}
%then (\ref{fqeqs}) becomes
%\begin{align}
%\label{FGrel}
%\lambda(\lambda-1-\vep) F_{\lambda 0}(Z) = 
%\bigg ( Z {\d \over \d Z} - 1-\vep \bigg )  G_{\lambda 0}(Z) \, ,
%\end{align}
%which has the solution
%\begin{align}
%\label{GZsol}
%G_{\lambda 0} (Z) = \lambda \, (-4Z)^{\lambda}
%{}_3F_2\bigg ({\lambda-1-\vep,\lambda+b,\lambda-b\atop 2\lambda-\vep, 
%\lambda+\half} ; Z \bigg ) \, .
%\end{align}

The recursion relations  (\ref{recurf}) for $a=0$ can be applied to 
obtain corresponding relations for $F_{\lambda\, n}$ relating even and
odd $n$. From (\ref{recurfa}) for $\ell = 2n$ we obtain
\begin{align}
\label{recurZh}
& {2\lambda - 2 n - 1 - 2 \vep \over 2\lambda - 2 n - 1 - \vep} \, (2n+\vep) \,
\big (  \lambda- n - 1 - \vep  + \delta_Z \big ) F_{\lambda\, 2n }(Z)\nn \\
&{}= (2\lambda - 4 n - 1 - 2 \vep ) (n+\vep) \, F_{\lambda\, 2n+1}(Z) + 
(2\lambda-1) n\,  F_{\lambda-1\, 2n-1}(Z) \, .
\end{align}
Also we may obtain
\begin{align}
\label{recurZg}
& \big ( \half (2n-1)  + \delta_Z \big ) F_{\lambda\, 2n }(Z) \nn \\
&{}= \half (2\lambda-1) \, F_{\lambda-1\, 2n-1}(Z) \nn \\
& \quad - {(2\lambda- 2n -1)(\lambda-n-\vep)\over
(2\lambda-2n-1 -\vep)(2\lambda-2n-\vep)}\,
{(\lambda-2n-\vep)^2-b^2 \over 4(2\lambda-4n+1-2\vep)} \,F_{\lambda\, 2n-1}(Z) \, ,
\end{align}
which with (\ref{recurZh})\ is sufficient to determine 
$F_{\lambda\, n }$ for all $n$ starting from $F_{\lambda\, 0}$ which is given
by \eqref{Fone}

To solve (\ref{recurZh})\ and (\ref{recurZg})\ it is convenient to adopt a Mellin transform
representation, adapting the result
for ${}_3F_2$ functions. We therefore assume
\begin{subequations}
\label{Mellin}
\begin{align}
\label{Mellg}
\hspace{-4mm}F_{\lambda\, 2n }(Z) = {}& 4^{\lambda -n} \, {\Gamma(2\lambda-2n-\vep) \, 
\Gamma(\lambda+\half)\over \Gamma(\lambda-n-\vep)\, \Gamma(\lambda-n +b) \,
\Gamma(\lambda-n-b)} \nn \\
\hspace{-4mm}&{}\times  {1\over 2\pi i} \int \! \d s \;
{\Gamma(s-\vep) \, \Gamma(s+b) \, \Gamma(s-b) \over 
\Gamma(s+ \lambda -n - \vep ) \, \Gamma(s + n+ \half)} \, 
\Gamma(-s+\lambda-n) \,
\alpha_{\lambda\, n }(s) \, (-Z)^s \, ,
\end{align}
and
\begin{align}
\label{Mellh}
\hspace{-4mm} F_{\lambda\, 2n+1 }(Z) = {} 4^{\lambda -n} \, {\Gamma(2\lambda-2n-1-\vep) \,
\Gamma(\lambda+\half)\over \Gamma(\lambda-n-\vep)\, \Gamma(\lambda-n +b) \,
\Gamma(\lambda-n-b)}& \nn \\ 
\hspace{-8mm}{}\times  {1\over 2\pi i} \int \! \d s \;
{\Gamma(s-\vep) \, \Gamma(s+b) \, \Gamma(s-b) \over 
\Gamma(s+ \lambda -n -1- \vep) \, \Gamma(s + n+ \half)} \, 
&\Gamma(-s+\lambda-n) \, \beta_{\lambda\, n }(s) \, (-Z)^s \, ,
\end{align}
\end{subequations}
The contour of the $s$-integration is parallel to the imaginary axis between 
the poles of $\Gamma(-s+\lambda-n)$ and those generated by the other 
$\Gamma$ functions,
moving the contour to the right gives a series expansion in powers
$(-Z)^{\lambda-n+p}$, $p=0,1,2\dots$.
For $n=0$ the result (\ref{Fone})\ requires
\begin{align}
\label{start}
\alpha_{\lambda\, 0}(s) = 1 \, .
\end{align}
The recursion relations (\ref{recurZh})\ and (\ref{recurZg})\ give
\begin{align}
\label{recura}
& (\lambda-2n-\half-\vep)(n+\vep) \, \beta_{\lambda\, n }(s) \cr
&\ {} = (\lambda-n-\half-\vep)(2n+\vep) \, \alpha_{\lambda\, n }(s)
- n \, (s+n-\half) \, \beta_{\lambda-1\, n-1 }(s) \, ,
\end{align}
and
\begin{align}
\label{recurb}
& (2\lambda-2n-1-\vep) \, \alpha_{\lambda\, n }(s) \cr
&\ {} = (s+ \lambda - n- 1 - \vep ) \, \beta_{\lambda-1\, n-1 }(s) \cr
& \qquad {}+ (s - \lambda + n) \, {\lambda- n -\half\over 
\lambda-2n+\half -\vep} \,
{(\lambda-2n-\vep)^2-b^2 \over (\lambda-n+b)(\lambda-n-b)} 
\, \beta_{\lambda\, n-1 }(s) \, .
\end{align}
Assuming (\ref{start}), (\ref{recura})\ and (\ref{recurb})\ then determines $\alpha_{\lambda\, n}(s)$
and $\beta_{\lambda\, n}(s)$ for all $n$ as polynomials in $s$ of
degree $n$. Clearly from (\ref{recura})\ $\beta_{\lambda\, 0}(s) = 1$ 
so that $F_{\lambda\, 1}(Z)$, just as $F_{\lambda\, 0}(Z)$, is given by a 
single ${}_3 F_2$ function. This generalises eq.~(4.11) of \cite{ElShowk:2012ht} to the case of general $b$. In general
\begin{align}
\label{norm}
 \alpha_{\lambda\, n }(\lambda-n) =  \beta_{\lambda\, n }(\lambda-n) = 1 \, ,
\end{align}
as is necessary to ensure (\ref{fZlim}).

Expressions for $ \alpha_{\lambda\, n }(s),\beta_{\lambda\, n }(s)$ for
general $n$ are rather convoluted. Expressions which can be extended
to arbitrary $n$ can be obtained by considering an expansion in polynomials
which can be absorbed into the $\Gamma$ functions in \reef{Mellin}. 
Then the individual terms in the Mellin integral correspond to ${}_3F_2$ 
functions with shifted arguments. 
By trial and error, a nice basis is provided by the polynomials
\beq
(s+b)_r(s+\half)_{n}/(s+\half)_{r},\qquad r=0,\ldots, n\,.\\
\eeq 
This ensures $\Gamma(s+b) \to \Gamma(s+b+r)$ and ${\Gamma(s+n+\half)^{-1}\to \Gamma(s+r+\half)^{-1}}$ in the integrand of \reef{Mellin}.
Iterating (\ref{recura}), (\ref{recurb}) and expanding in this basis, all the coefficients factorise nicely 
and the results for  $ \alpha_{\lambda\, n }(s),\beta_{\lambda\, n }(s)$ become
\begin{align}
\label{ansol}
\alpha_{\lambda\, n }(s) &=
{(s+\half)_n \, (\lambda-\vep-b -2n)_n \over (\lambda-b-n)_n \, 
(\lambda -\vep -2n+\half)_n}\nn\\ 
&\hspace{3cm}\times {}_4F_3\bigg ({-n,s+b,b+\half,\vep+n\atop
s+\half,-\lambda+\vep+b+n+1, \lambda+b-n} ; 1 \bigg ) \, ,\\
\beta_{\lambda\, n }(s) &= 
{(s+\half)_n \, (\lambda-\vep-b -2n-1)_n \over (\lambda-b-n)_n \,
(\lambda -\vep -2n-\half)_n}\,\nn\\
&\hspace{3cm}\times {}_4F_3\bigg ({-n,s+b,b+\half,\vep+n+1\atop
s+\half,-\lambda+\vep+b+n+2, \lambda+b-n} ; 1 \bigg ) \, .
\end{align}

Terminating ${}_4F_3(1)$ functions satisfy Whipple identities
\begin{align}
\label{Whip}
{}_4F_3\bigg ( {-n,a,b,c \atop d,e,f};1 & \bigg ) =
{(e-a)_n(f-a)_n \over (e)_n\, (f)_n} \,
{}_4F_3\bigg ({-n,a,d-b,d-c \atop d,a-e+1-n,a-f+1-n};1 \bigg ) \, ,
\end{align}
so long as they are balanced, i.e.~if
\begin{align}
\label{balance}
a+b+c-n+1 = d+e+f \, .
\end{align}
This condition is satisfied in (\ref{ansol}).
The relation (\ref{Whip}), together with various equivalent permutations, 
then allows (\ref{ansol})\ to be written in a variety of 
alternative forms, such as demonstrating that the results are invariant 
under $b\to -b$. Balanced terminating ${}_4F_3(1)$ functions also 
satisfy three term linear contiguous relations,
where two of the parameters $-n,a,b,c,d,e,f$ are shifted by 
$\pm 1$ while maintaining the 
balancing condition in (\ref{balance}) \cite{Andrews}. These are sufficient to show that
(\ref{ansol})\ satisfies (\ref{recura})\ and (\ref{recurb}).

With these results in (\ref{Mellin}), the expression (\ref{Fone})\ generalises to a 
finite sum of ${}_3F_2$ functions
\begin{align}
\label{resgh}
F_{\lambda \, 2n} (Z) = {}& 
(-4Z)^{\lambda-n} {(\lambda-n+\half)_n \over (\lambda-n- b)_n \,
(\lambda-\vep-2n + \half)_n} \cr
&{}\times \sum_{r=0}^n {n \choose r} \,
{(b+\half)_r \, (\vep+n)_r \, (\lambda -\vep -b -2n)_{n-r} \over
(\lambda - n + \half )_r } \cr
& \hskip 1.5cm {}\times
{}_3F_2\bigg ({\lambda-n-\vep,\lambda-n-b,\lambda-n +b+r\atop 
2\lambda-2n-\vep, \lambda-n+r+\half} ; Z \bigg ) \, ,\cr
F_{\lambda \, 2n+1} (Z) = {}&
(-4Z)^{\lambda-n} {(\lambda-n+\half)_n \over (\lambda-n- b)_n \,
(\lambda-\vep-2n - \half)_n} \cr
&{}\times \sum_{r=0}^n {n \choose r} \,
{(b+\half)_r \, (\vep+n+1)_r \, (\lambda -\vep -b -2n-1)_{n-r} \over
(\lambda - n + \half )_r } \cr
& \hskip 1.5cm {}\times
{}_3F_2\bigg ({\lambda-n-\vep,\lambda-n-b,\lambda-n +b+r\atop
2\lambda-2n-1-\vep, \lambda-n+r+\half} ; Z \bigg ) \, .
\end{align}
These expressions for higher spin blocks are more concise than 
those used in Ref.~\cite{ElShowk:2012ht}. Using eq.~(4.9) of \cite{ElShowk:2012ht} 
the diagonal limit of the spin $\l\ge 2$ conformal block is expressed as a 
sum of $\l+1$ terms of ${}_3F_2$ type which is essentially twice
the number appearing in  \eqref{resgh}.

\section{Computing conformal blocks and their derivatives efficiently}
\label{sec:algo}

We will now return to the case of general $a,b$, possibly both of them nonzero. 
In this section we will present an efficient algorithm to compute conformal 
blocks in the derivative expansion around any point $z=\bar z=t_0$ 
on the diagonal. Such expansions form the basic input for the numerical 
conformal bootstrap algorithms, where $t_0=\frac{1}{2}$ is normally used. 
Our algorithm is an extension of the method first used in 
\cite{ElShowk:2012ht}, with several improvements made possible by 
the results of \cite{Hogervorst:2013sma} and of the present paper. 

Let's denote by $h_{m,n}$ conformal blocks derivatives with respect 
to the coordinates $t,s$ (related by simple rescalings to $a,b$ in \cite{ElShowk:2012ht}):
\begin{align}
&F_{\lambda_1 \lambda_2}(z,\bar z)=\sum_{m,n\ge0}\,\frac{1}{m!n!}\,   h_{m,n}\, (t-t_0)^m s^n \,,
\end{align}
where
\beq
z=t+\sqrt{s},\quad \bar z=t-\sqrt{s}\,.
\eeq
The diagonal corresponds to $s=0$ and \eqref{eq:neardiag} corresponds to
keeping just $n=0,1$.
Conformal blocks being symmetric in $z\leftrightarrow \bar z$, ensures that 
the expansion is in integer powers of $s$.

The first observation is that the derivatives in the direction orthogonal 
to the diagonal ($s$-derivatives) can be recursively determined from 
the derivatives along the diagonal. This recursion follows from the 
quadratic Casimir equation in \reef{eq:24} and has the following schematic structure: 
\begin{align}
\label{eq:CKshort}
h_{m,n}=\sum_{m'\le m-1} (\ldots) h_{m',n} +\sum_{m'\le m+2}\left[ (\ldots) h_{m',n-1} + (n-1) (\ldots) h_{m',n-2} \right]\,.
\end{align}
The precise coefficients for $a,b=0$ and for $t_0=\tfrac{1}{2}$ were given in \cite{ElShowk:2012ht}; extension to $a,b$ nonzero and general $0<t_0<1$ is straightforward. 
By this recursion, moving one unit up in $n$ we lose two units in $m$. So knowing the derivatives $h_{m,0}$ for $m=0,\ldots,m_{\rm max}$ is sufficient to compute $h_{m,n}$ for all $m+2n\le m_{\rm max}$. In practical applications of numerical bootstrap, it is common to use the derivatives in such a triangular table with $m_{\rm max}$ up to ${\rm O}(20)$ or more.

Hence, we are reduced to computing the derivatives $h_m\equiv h_{m,0}$ along the diagonal. We next observe that $h_m$ satisfy another set of recursion relations as a consequence of the ODEs \reef{eq:ODEs}. These recursions have the schematic form:
\begin{align}
\label{eq:hm}
m(m-1)(m-2)(m-3) &h_{m}=\sum_{m'=\min(0,m-7)}^{m-1} (\ldots) h_{m'}\,\quad \text{($\l=1,2,\dots $)}\,,\\
m(m-1)(m-2) &h_{m}=\sum_{m'=\min(0,m-5)}^{m-1} (\ldots) h_{m'}\,\quad \text{($\l=0$)}\,.
\end{align}
The coefficients follow trivially from \reef{eq:ODEs} so we do not give them 
here. Assuming that the first few $h_m$ are known (namely for $m=0,1,2,3$ 
for general $\l$ and $m=0,1,2$ for $\l=0$), the rest can be found by these recursions.

Thus, it remains to find a method to compute the derivatives $h_m$ at low $m$. This can be done as follows. Conformal blocks on the diagonal have an expansion \reef{eq:seriesf} around $z=0$. As we explained in section \ref{sec:diffeq}, the expansion coefficients $a_n$ are fixed by the ODEs \reef{eq:ODEs}. The closed-form expressions for $a_n$ are not available, but they can be found up to an arbitrary order via the recursion relations given in appendix \ref{sec:recursions}. Crucially, since $z=0$ is a regular singular point, the single normalisation condition $a_0=1$ is sufficient to determine all of $a_n$. This is unlike the recursions \reef{eq:hm} for derivatives $h_n$ around a regular point $0<t_0<1$ where derivatives up to the equation order minus one have to be supplied as the initial condition. 

A natural method to compute the $h_m$ at low $m$ is then just to evaluate $a_n$ up to a sufficiently high order $N$, 
and to sum up the series \reef{eq:seriesf} by differentiating term by term:
\beq
\label{eq:hm2}
h_m=(\del_z)^m f_{\lambda_1\lambda_2}(z)|_{z=t_0}\approx 
\sum_{n=0}^{N} a_n(\Delta+n)(\Delta+n-1)\ldots(\Delta+n-m)\,  t_0^{\Delta+n-m} \,.
\eeq
% MH: the label eq:hm is defined before as well 
We propose to use this method but with a small modification, which greatly improves its numerical efficiency. 
Namely, we will evaluate the conformal blocks and their derivatives expanding not in $z$ but in the variable $\rho$ related to $z$ by
\begin{align}
\rho=\frac{z}{(1+\sqrt{1-z})^2},\qquad z=\frac{4\rho}{(1+\rho)^2}\,.
\end{align}
For $\rho$ the points $0,1,\infty$ are mapped to $0,1,-1$ and $z\to z'$
corresponds to $\rho\to - \rho$.
The corresponding expansion coefficients will be denoted $b_n$:
\beq
\label{eq:frho}
f_{\lambda_1\lambda_2}(\rho)= (4\rho)^\Delta \sum_{n=0}^\infty b_n\rho^n,\qquad b_0= 1 \,.
\eeq
The variable $\rho$ was introduced in \cite{Pappadopulo:2012jk} as the coordinate corresponding to the OPE frame where the operators are inserted symmetrically with respect to the radial quantisation origin. Ref.~\cite{Hogervorst:2013sma} explained that $\rho$ is a much better in expansion parameter for the conformal blocks than $z$, for two reasons. First, $\rho$ is smaller that $z$
 (e.g.~$\rho(\tfrac{1}{2})\approx 0.17$) for the ranges of physical interest and in applications 
 the series \reef{eq:frho} converges faster. Second, the coefficients $a_n$ grow as $\Delta^n$ for large $\Delta$, while $b_n$ remain bounded in this limit.
So for large $\Delta$ more and more terms will have to be retained in the $z$-series, while the $\rho$-series will not suffer from this drawback.

The ODEs \reef{eq:ODEs} in the variable $\rho$ take the following form:
\beq
\label{eq:Drho4} \calD_{4} f_{\la_1 \la_2}(\rho) = 0, \qquad
 \calD_3 f_{\la\,  \la}(\rho) = 0, 
\eeq
where
\begin{subequations}
\begin{align}\calD_4 ={}& (\rho -1)^3 \rho ^4 (\rho +1)^4 \diffop{\rho}{4} + 
  2 (\rho -1)^2 \rho ^3 (\rho +1)^3 \big\{(2 \vareps +5) \rho ^2+8 S \rho +2 \vareps -1\big\} \diffop{\rho}{3}  \nn \\
& -2 (\rho -1) \rho ^2 (\rho +1)^2 \Big\{[ c_2-(\vareps +4) (2 \vareps +3)] \rho ^4+4 [P-3 S (2 \vareps +3)] \rho ^3  \nn \\ 
\noalign{\vskip -2pt}
& \hskip 1.2cm -2 \left[20 S^2+2 \vareps ^2+ c_2+4 P+3 \vareps -5\right] \rho ^2+4 [P+S (3-6 \vareps )] \rho -2 \vareps ^2+ c_2+\vareps\Big\} \diffop{\rho}{2} \nn \\  
& -2 \rho  (\rho +1) \Big\{(2 \vareps +3) [ c_2-2 (\vareps +1)] \rho ^6+[12 \vareps  P+6 P+8  c_2 S-8 S (\vareps +1) (2 \vareps +3)] \rho ^5 \nn \\
&  \hskip 2cm+ \left[4 \{-4 (4 \vareps +3) S^2-2 \vareps ^2+8 P (S-\vareps )+\vareps +3\}- c_2 (2 \vareps +5)\right] \rho ^4 \nn \\
&  \hskip 2cm -4 \left[P (16 S-10 \vareps +5)+2 S \left(8 S^2+4 \vareps ^2+2  c_2-5\right)\right] \rho ^3 \nn \\
&  \hskip 2cm + \left[-2 \vareps   c_2+ c_2+16 P (2 S-2 \vareps +1)-2 \left(16 S^2+\vareps -1\right) (2 \vareps -1)\right] \rho ^2 \nn \\
&  \hskip 2cm +\left[2 P (6 \vareps -1)+8 S \left(-2 \vareps ^2+\vareps + c_2\right)\right] \rho + c_2+2  c_2 \vareps\Big\} \frac{\mrm{d}}{\mrm{d}\rho}  \displaybreak[0] \nn \\
& + (1-\rho) \big\{[ c_2 (4 \vareps +2)- c_4] \rho ^6+2 [- c_4+2  c_2 (2 S+1) (2 \vareps +1)+4 P \vareps  (2 \vareps +1)] \rho ^5 \nn \\
&\qquad +\left[ c_4-16 P (-6 \vareps  S+S+2 \vareps^2 -3\vareps)+2  c_2 \left(16 S^2+ 16S \vareps +8S +6 \vareps -1\right)\right] \rho ^4 \nn\\
&\qquad +4 \left[ c_4+2  c_2 (2 S+1) (4 S+2 \vareps -1)+4 P \{8 S^2+(6-4 \vareps ) S+\vareps  (2 \vareps -3)\}\right] \rho ^3\nn\\
&\qquad +\left[ c_4-16 P \{-6 \vareps  S+S+\vareps  (2 \vareps -3)\}+2  c_2 \{16 S^2+8 (2 \vareps +1) S+6 \vareps -1\}\right] \rho ^2\nn\\
&\qquad +2 [- c_4+2  c_2 (2 S+1) (2 \vareps +1)+4 P \vareps  (2 \vareps +1)] \rho - c_4+ c_2 (4 \vareps +2)\big\} \, ,
\end{align}
and
\begin{align} \calD_3 ={}& (\rho -1)^2 \rho ^3 (\rho +1)^3 \diffop{\rho}{3} + 2 (\rho -1) \rho ^2 (\rho +1)^2 \big\{(\vareps +3) \rho ^2+6 S \rho +\vareps\big\}  \diffop{\rho}{2} \nn \\
& -2 \rho  (\rho +1) \Big\{( c_2-2 \vareps -3) \rho ^4 - 4S (2 \vareps +3) \rho ^3 \nn \\
\noalign{\vskip -4pt}
&\hskip 2.5cm -\left[2  c_2 + 2 (8 S^2+\vareps ) -1\right] \rho ^2 -8 S \vareps\, \rho + c_2 + 4P \rho(\rho-1)^2 \Big\}  \frac{\mrm{d}}{\mrm{d}\rho} \nn \\
%%% corrected:
% & -2 (\rho -1) \Big\{ c_2 \rho ^4+2 [(1+2 S)  c_2+(1+2\vareps) P  ] (\rho^3+\rho)\\
 % &\hskip 5cm +2 [(1+4 S)  c_2+2 P (4 S-2 \vareps +1)] \rho ^2 + c_2\Big\}\,. \\
 %%% latest version:
& -2 (\rho -1) \Big\{ \left[ c_2 (\rho^2+1) + 4 S c_2 \rho \right](\rho + 1)^2  \nn \\
& \hskip 4.0cm + 2 P (1+2\vareps) (\rho^3 + \rho) + 4 P (4S-2\vareps+1)\rho^2 \Big\} \,. 
\end{align}
\end{subequations}

They imply recursion relations which determine all $b_n$ starting from $b_0=1$. We do 
not present these recursions here, since they are totally analogous to those for 
the $a_n$ given and analysed in appendix \ref{sec:recursions}.

Our method for evaluating $h_m$ for low $m$ is thus as follows. First evaluate $b_n$ up to a sufficiently high order $N$ so that the series 
\beq
\label{eq:delrho}
(\del_\rho)^mf_{\lambda_1\lambda_2}(\rho)|_{\rho=\rho_0}\approx \sum_{n=0}^{N} b_n(\Delta+n)(\Delta+n-1)\ldots(\Delta+n-m) \rho_0^{\Delta+n-m} \,,\quad \rho_0\equiv\rho(t_0),
\eeq
give a good approximation to the RHS for all $0\le m\le m_0$, where $m_0$ is the maximal needed derivative order. The accuracy of this approximation can be controlled via the asymptotics of the $b_n$ coefficients, which can be understood from the recursion relations that they satisfy. For the reasons given above, the needed number of terms $N$ in this series will be much smaller than in \reef{eq:hm}.
Via the inverse change of variables $\rho\to z$, the derivatives in $z$ can then be expressed as linear combinations of derivatives in $\rho$.

\section{Summary}
\label{sec:summary}

In this paper we have hopefully made several  additions to the theory of conformal blocks
for general dimension $d$ which should be of assistance in  implementations of
the conformal bootstrap:
\begin{enumerate}
\item We derived ODEs satisfied by the blocks on the diagonal $z=\bar z$. Such equations were known before
only for spins $\l=0,1$ and only for the case of equal external scalar dimensions $\Delta_1=\Delta_2$, $\Delta_3=\Delta_4$ \cite{ElShowk:2012ht}.
\item We proposed an algorithm for an efficient numerical evaluation of conformal blocks and their derivatives around any point on the diagonal,
generalising the method first used in \cite{ElShowk:2012ht} for the case of equal external scalar dimensions.

\item For the partial case $\Delta_1=\Delta_2$ we derived closed-form representations for the conformal blocks on the diagonal in terms of ${}_3F_2$ functions. Such representations were known before only for $\l=0$ \cite{ElShowk:2012ht,Paulos-notes}, and for $\l\ge1$ for the case when the second pair of dimensions are also equal: $\Delta_3=\Delta_4$ \cite{ElShowk:2012ht}. For $\l>1$ our representations are more compact and explicit than those in \cite{ElShowk:2012ht}.
\end{enumerate}

The numerical algorithm from section \ref{sec:algo} has an immediate application. Starting from \cite{Rattazzi:2008pe} and until now, all numerical conformal bootstrap studies have focused on correlators with equal external dimensions. As explained in \cite{ElShowk:2012ht} in the context of the 3$d$ Ising model, it would be interesting to perform simultaneous analysis of several four point functions, e.g.~$\langle \sigma\sigma\sigma\sigma\rangle$, 
$\langle \sigma\sigma\eps\eps\rangle$, and $\langle \eps\eps\eps\eps\rangle$, where $\sigma$ and $\eps$ are the lowest-dimension $\bZ_2$-odd and -even operators. Conformal blocks needed for the $\langle \sigma\sigma\eps\eps\rangle$ correlator involve unequal external dimensions in two out of three OPE channels. They can be computed using our algorithm.

From a more mathematical perspective the ordinary differential equations obtained here may
allow more insight into that analytic structure of conformal blocks in general dimensions and
$\ell>0$. When conformal blocks are expressible in terms of ordinary hypergeometric functions,
as in even dimensions, their monodromy can be derived from the classic 19th century results
of Riemann. The extension to ${}_n F_{n-1}$ functions, as appear in some of our solutions,
was done much later \cite{beukers1989monodromy} and such techniques may be extendible to the 4th- 
and 3rd-order differential operators obtained here. 

\section*{Acknowledgements}
The work of S.R. is supported in part by the European Program ``Unification in the LHC Era", 
contract PITN-GA-2009-237920 (UNILHC). H.O would like to remember his lengthy 
collaboration with Francis Dolan on conformal blocks and also to thank Damian
Reding for helpful mathematical advice.

\appendix

\section{Factorisation}
\label{sec:fact}
Differential equations may in some circumstances be reduced to
ones of lower degree. Such simplifications are in general a non-trivial
mathematical problem \cite{bobienski2011reduction}.
Here we consider some cases when the
4th-and 3rd-order differential operators factorise. In general 
differential operators of the form \eqref{eq:Dnj}, subject to \eqref{eq:cases}, 
may factorise so that
\beq
D^{(n'+n,j'+j)} = D^{(n',j')} D^{(n,j)} \, .
\eeq
If this is possible the exponents for $ D^{(n,j)} $ must be a subset of those
for $D^{(n'+n,j'+j)} $ satisfying \eqref{Frob}. In this case solutions of $D^{(n'+n,j'+j)}u=0$
satisfy a lower order differential equation if their exponents correspond to
those for $D^{(n,j)}$. We discuss in this appendix some cases
when this happens for the operators of interest here.

First we note that the fourth-order operator contains the third-order one
when $\ell=0$,
\beq
D_z^{(4,3)}\big |_{\lambda_1=\lambda_2=\lambda} = 
\bigg ( (z-1) z \frac{\d}{\d z} + (a+b-1-\vareps ) z + 2\vareps+ 1 \bigg ) D_z^{(3,2)} \, ,
\eeq
where $D_z^{(4,3)}, D_z^{(3,2)}$ are determined by \eqref{D4p}, \eqref{D3p} respectively.

There is also a simplification for conformal blocks with $\lambda_2=\vep-a$
as a consequence of 
\beq
D_z^{(4,3)}\big |_{\lambda_1=\lambda, \lambda_2= \vareps - a } = 
D_z^{(2,2)}\,  D_z^{(2,1)} \, ,
\eeq
where
$ D_z^{(2,1)}$ is an operator with exponents
\beq
\text{
\begin{tabular}{  c   c   c   }
0 &$ \infty$& 1 \\
\hline
$\lambda+\vareps-a$ & $2a$ & 0 \\ 
$1-\lambda+\vareps-a$& $a+b-\vareps $ &$ -a-b-\vareps$ 
\end{tabular}
}
\eeq

\noindent
The form of $D_z^{(2,2)}$ is unimportant but
\begin{align}
D_z^{(2,1)} = {}& (z-1) z^2 {{\rm d}^2 \over {\rm d } z^2} +
 \big ( (3a+b+1-\vareps) z + 2(\vareps-a) \big )z  {{\rm d} \over {\rm d } z} \nn \\
&{} + 2a(a+b-\vareps) z +(\lambda+\vareps -a) (\lambda-\vareps +a -1) \, ,
\end{align}
so that the solution for the diagonal conformal blocks, with $\lambda=\ell+\vareps-a$,
becomes an ordinary hypergeometric function
\beq
f_{\lambda \, \vareps-a}(z) = z^{\ell+2\vareps-2a} 
F(\ell+\vareps+b-a,\ell+2\vareps;2\ell + 2 \vareps -2a;z) \, .
\label{twist}
\eeq
For $a=b=0$ this result was obtained in \cite{Hogervorst:2013sma}. For $a=0$ 
these blocks correspond to the ``leading twist" operators which saturate the unitarity bound for $\l\ge 1$, while $a>0$ is a suitable generalisation. 

Two other special cases arise for $d=1,3$ and $\ell=0$  when the operator
given by \eqref{eq:Dnj} and \eqref{D3p} factors as
\begin{align}
&D_z^{(3,2)}\big |_{\vep =- {1\over 2}}
 = - \Big ( (z-1)z {\d \over \d z} + (a+b-\half )z + 1 \Big ) 
 \Big( D_z(2a,2b) - 4\lambda (\lambda- \half )\Big)\,,\\
&D_z^{(3,2)}\big |_{\vep ={1\over 2}} 
= z\Big (D_z(2a,2b)-4(\lambda-\half)(\lambda-1)\Big)
\frac 1z
\Big (z(1-z)\diffop{z}{1}-(a+b-\half)z-1\Big)\,.
\end{align}
Hence, for $d=1$ the solutions with the right asymptotics are of the form $z^{2\lambda}$ times an ordinary hypergeometric function \cite{DO3},
while for $d=3$ they can be written as integrals thereof:
\beq
f_{\lambda\lambda}(z)\big |_{\vep ={1\over 2}} = 
\frac {(2\lambda-1)z}{(1-z)^{a+b+1/2}} \int_0^z {\rm d}t\,(1-t)^{a+b-1/2}\, t^{2\lambda-2}
F(2\lambda-1+2a,2\lambda-1+2b;4\lambda-2;t)\,.
\eeq

\section{Recursion relations for $a_n$}
\label{sec:recursions}

Via Frobenius' method, the ODEs~(\ref{eq:ODEs}) imply recursion relations for the 
coefficients $a_n$ in \reef{eq:seriesf}, four-term for general $\l$: 
\beq
\label{eq:recursEq}
n (\DD+\ell+n -1) (\DD-\ell+n -2 \vareps -1) (2\DD+n -2 \vareps -2) a_n  = \sum_{i=1}^3 \gamma_{i,n} \, a_{n-i},\\
\eeq
where
\begin{align*}
\gamma_{1,n} &= 3 n^4+\big(4 S+12 \DD -10 \vareps -19\big) \, n^3  \\
& \quad +\big[8 \vareps ^2-2 (6 S+15 \DD -22) \vareps -4  {c_2}+2 P+3 (2 \DD -3) (2 S+3 \DD -5)\big] n^2 \, \\
& \quad +\big[12 \DD ^3+12 S \DD ^2-57 \DD ^2-36 S \DD +90 \DD +4 (2 S+4 \DD -5) \vareps ^2+26 S \, \\
& \qquad +P (4 \DD -6 \vareps -5)-2 (3 \DD -4) (4 S+5 \DD -8) \vareps + {c_2} (-4 S-8 \DD +6 \vareps +13)-47\big] n \, \\
& \quad +\big  [\DD  (3 \DD -4 \vareps -10)+S (4 \DD -4 \vareps -6)+6 \vareps +9\big ] (\DD -1)(\DD -2 \vareps -2)\\
&\quad +c_4 +P (\DD -2 \vareps -1) (2 \DD -2 \vareps -3)\\
&\quad +{c_2} \big [-10 \vareps +S (-4 \DD +4 \vareps +6)+\DD  (-4 \DD +6 \vareps +13)-11\big ],\\[5pt]
\gamma_{2,n} &= (\DD +n+S-\vareps -2) \big\{-\!3 n ^3+\big[-5 S-9 \DD +5 (\vareps +4)\big] n^2\, \\
&\quad +\big[2  {c_2}-4 P-3 (7 \vareps +15)+S (-10 \DD +8 \vareps +21)+\DD  (10 \vareps +40-9 \DD )\big] n\, \\
&\quad +{c_2} (2 S+2 \DD -5)+P (-4 \DD +6 \vareps +7)\,\\
&\qquad -(\DD -2)\big  [S (5 \DD -8 \vareps -11)+\DD  (3 \DD -5 \vareps -14)+11 \vareps +17\big ]\,\big\}, \,\\[5pt]
\gamma_{3,n} &= (\DD+2 a +n-3) (\DD+2 b +n-3) (\DD +n+S-\vareps -3) (\DD +n+S-\vareps -2),
\end{align*}
and three-term for $\l=0$:
\beq
n (\DD +n-1) (2 \DD +n-2 \vareps -2) \, a_n  = \sum_{i=1}^2 \xi_{i,n} \, a_{n-i} \, ,
\label{eq:recursEq0}
\eeq
for 
\begin{align}
\xi_{1,n} = {}& 2 n^3+3 \big(S+2 \DD -\vareps -3\big) \, n^2\nn\\
&{}+\big[2 P+S (6 \DD -4 \vareps -9)+7 \vareps +\DD  (5 \DD -4 \vareps -16) +13\big] n\nn\\
&{} +\half  \big [2 P+(\DD -2) (2 S+\DD -2)\big ] (2 \DD -2 \vareps -3)\, , \nn\\[5pt]
\xi_{2,n} = {}& - (\DD+2 a+n -2) (\DD+2 b+n -2) (\DD+S+n -\vareps -2)\,  .\nn
\end{align}

Apart from a few cases listed below, these recursion relations determine $a_n$ for $n\ge 1$ starting from $a_0=1$
and imposing $a_n=0$ for $n<0$. 
The exceptions occur when the factor multiplying $a_n$ in the LHS vanishes for some $n=n_0\ge 1$. These are the cases when $\Delta$ is smaller by the positive integer, $n_0$, than the largest characteristic exponent at $z=0$ in \reef{eq:exponents}. Frobenius' method requires a special treatment in such situations. As we will see below, in our case potential ambiguities can be resolved using as an extra physical input the fact that conformal blocks should be  continuous functions of $\Delta$.

There are actually only three exceptional cases consistent with the unitarity 
bounds and assuming $d\ge 2$. We consider them one by one.

\noindent
1. $\l>0,\DD = \l + 2\vareps,n_0=1$ in \reef{eq:recursEq}

This corresponds to a spin $\l>0$ primary higher spin conserved current
saturating the unitarity bound. Conformal block of such a primary are 
defined only for $a=b=0$, since three point functions with scalars of 
unequal dimension vanish by imposing conservation. 
Thus we should not worry that eq.~\reef{eq:recursEq} predicts $a_1\to \infty$ for $a$ or $b$ nonzero and $\DD \to\l + 2\vareps$.
However for $a=b=0$ conformal blocks 
should be continuous in this limit. 

Indeed, for $n=1$ and $a=b=0$, $\gamma_{1,1}$ factorises:
\beq 
\gamma_{1,1} = \tfrac{1}{2}(\DD-\l - 2\vareps)(2\DD-2\vareps-1)\DD(\DD+\ell)\,.
\eeq
The offending factor $(\DD-\l - 2\vareps)$ now appears on both sides of the recursion relation defining $a_1$. Cancelling this factor
gives an equation for $a_1$ which is continuous in the full range including the unitarity bound. Once $a_1$ is computed according to this prescription, the rest of the coefficients follow from the recursion relation unambiguously.

We note in passing that the diagonal limit of spin $\l$ blocks saturating the unitarity bound is known in closed form for arbitrary $d$ \cite{Hogervorst:2013sma}; see \eqref{twist} for $a=b=0$.

\noindent
2. $\l=0, \DD = \vareps + \half,n_0=1$ in \reef{eq:recursEq0}

Again, these conformal blocks can and should be defined by continuity. 
Namely, one can check that for $n=1$ and any $a$ and $b$ the RHS of \reef{eq:recursEq0} factorises, so that the problematic factor $(2 \DD -2 \vareps -1)$ can be cancelled from both sides of the recursion relation. Once $a_1$ is defined this way, the rest of the coefficients follow from the recursion.

\noindent
3. $\l=0, \DD = \vareps,n_0=2$ in \reef{eq:recursEq0}

This case corresponds to a scalar field at the unitarity bound, so it is not particularly interesting. This conformal block can occur only in the free scalar theory, and only if the external field dimensions differ by the free scalar dimension $\vep$. We have in mind the OPE 
$:\!\phi\,\calO\!:\!\times \, \calO\supset \phi$.
One can check that in this case, i.e.~for $a,b=\pm\vep/2$, the free scalar block can be defined by continuity exactly as above. Namely the RHS of \reef{eq:recursEq0} for $n=2$ can be factored (plugging in $a_1$ computed in the previous recursion step) and the offending factor cancelled. For all the other choices of $a,b$ the scalar block diverges when $\DD \to \vareps$.

\bibliographystyle{utphys}
\bibliography{Diag-biblio}

\end{document}